\documentclass{ws-procs10x7}

\def \ifb {fb$^{-1}$}

\def \lambdahhh {\ifmmode \lambda_{hhh}\else $\lambda_{hhh}$\fi}
\def \lambdahhhzero {\ifmmode \lambda_{hhh}^{(0)}\else
  $\lambda_{hhh}^{(0)}$\fi}
\def \mh {\ifmmode m_h \else $m_h$\fi}

\def    \=              {\;=\;} 
\def    \frac           #1#2{{#1 \over #2}}

\def    \lsim {\raisebox{-3pt}{$\>\stackrel{<}{\scriptstyle\sim}\>$}}  
\def    \gsim {\raisebox{-3pt}{$\>\stackrel{>}{\scriptstyle\sim}\>$}}

\newcommand     \be     {\begin{equation}} 
\newcommand     \ee     {\end{equation}} 
\newcommand     \ba     {\begin{eqnarray}} 
\newcommand     \ea     {\end{eqnarray}}

\newcommand     \ptmin     {\ifmmode p_{T}^{min} \else 
                            $p_{T}^{min}$ \fi}

\def \Emu  {\ifmmode{E_{\mu}
    }\else{$E_{\mu}$}\fi} 
\def \Enu  {\ifmmode{E_{\nu}
    }\else{$E_{\nu}$}\fi}

\def \nue  {\ifmmode{\nu_e}\else{$\nu_e$}\fi} 
\def \numu  {\ifmmode{\nu_{\mu}}\else{$\nu_{\mu}$}\fi}

\def \to   {\mbox{$\rightarrow$}}

\newcommand \jpsi{\ifmmode{J/\psi 
    }\else{$J/\psi$}\fi}

\def\Ord{\lower .7ex\hbox{$\;\stackrel{\textstyle <}{\sim}\;$}}
\def\OOrd{\lower .7ex\hbox{$\;\stackrel{\textstyle >}{\sim}\;$}}

\newcommand{\br}{\begin{eqnarray}}
\newcommand{\er}{\end{eqnarray}}
\newcommand{\bea}{\begin{eqnarray}}
\newcommand{\eea}{\end{eqnarray}}
\newcommand{\bn}{\begin{enumerate}}
\newcommand{\en}{\end{enumerate}}
\newcommand{\bc}{\begin{center}}
\newcommand{\ec}{\end{center}}

\def\epem{\ifmmode{e^+ e^-} \else{$e^+ e^-$} \fi}

\newcommand{\Dir}{\kern -6.4pt\Big{/}}
\newcommand{\Dirin}{\kern -10.4pt\Big{/}\kern 4.4pt}
\newcommand{\DDir}{\kern -7.6pt\Big{/}}
\newcommand{\DGir}{\kern -6.0pt\Big{/}}

\begin{document}

\title{Beyond the Standard Model Higgs Boson self-couplings at 
the LHC}
\author{M. Moretti}

\address{Dipartimento di Fisica -- Universit\`a di Ferrara and
INFN -- Sezione di Ferrara, Via Paradiso 12, 44100 Ferrara, 
Italy\\(Email: Mauro.Moretti@fe.infn.it)}

\author{S. Moretti}

\address{School of Physics \& Astronomy, University of Southampton,
Highfield, Southampton SO17 1BJ, UK\\(Email: Stefano@hep.phys.soton.ac.uk)}

\author{F. Piccinini}

\address{INFN -- Sezione di Pavia and Dipartimento di Fisica Nucleare 
e Teorica, Universit\`a di Pavia, via Bassi 6, 27100 Pavia,
Italy (Email: Fulvio.Piccinini@pv.infn.it)}

\author{R. Pittau}

\address{Departamento de F\'{\i}sica Te\'orica y del Cosmos
and Centro Andaluz de  F\'{\i}sica de Part\'{\i}culas 
Elementares (CAFPE), Universidad de Granada, E-18071 Granada, 
Spain (On leave from Dip. di Fisica 
Teorica and INFN, Universit\`a di Torino, Via Giuria 1, 10125 Torino,
Italy)\\(Email: Roberto.Pittau@to.infn.it)}

\author{A.D.~Polosa}

\address{Centro Studi 
e Ricerche ``E.~Fermi'', via Panisperna 89/A, 00184 Roma, 
Italy\\(Email: Antonio.Polosa@cern.ch)}

\twocolumn[\maketitle\abstract{{\tt Bari-TH 499/2004}, {\tt CAFPE-46/04}, 
{\tt FNT-T/2004/19}, {\tt UG-FT-176/04}, {\tt SHEP-04-07}
\\ 
\\
A study of two-Higgs-doublet models (2HDM) in the
decoupling limit reveals the existence of parameter configurations
with a large triple-Higgs self-coupling as the only low-energy trace of a
departure from a Standard Model (SM) Higgs sector. This observation
encourages attempts to search for double Higgs production at the Large
Hadron Collider (LHC) and its luminosity upgrade (SLHC) 
even in mass regions which have been shown to be
very hard to probe in the context of SM-like Higgs self-couplings. 
In a scenario where only an Intermediate Mass Higgs (IMH) 
boson, with 120~GeV~$\lsim~m_h~\lsim$~140 GeV, is discovered at the LHC, 
with measured couplings to fermions and gauge bosons compatible with 
their SM values, we show that Higgs-pair production (with each
Higgs state decaying in two 
$b\bar b$ pairs) through Weak Boson Fusion (WBF) could open a window on 
physics beyond the SM in the Higgs sector.}]

\section{Introduction}
\label{sec:intro}
Several detailed studies~\cite{AtlasCMS} have established
the ability of the ATLAS and CMS experiments to detect a SM
Higgs boson over the full range of allowed masses. Many
decay channels will be accessible at the LHC, enabling
the determination of several Higgs boson couplings
with accuracies that can be as good as 10\%~\cite{AtlasCMS}.
Possible significant departures from the SM expectations would 
allow to infer a non-standard Higgs, but there is no guarantee 
that non-SM Higgs sectors will become
manifest via these measurements. Examples are given by 
2HDMs, including Supersymmetric
models, where the spectrum of Higgs
bosons beyond the lightest one could be very heavy and the couplings
of the latter to fermions and gauge bosons reduce to those of the SM. 
In this contribution, we explore the possibility that, in this
{\it decoupling} scenario (namely only one SM-like Higgs boson $h$ 
discovered at LHC in the mass range $120-140$~GeV 
with couplings to fermions and 
gauge bosons compatible, within the foreseen 
experimental uncertainty, with the SM), only a large
deviation from the SM value $\lambda_{hhh}^{(0)}$
of the triple-Higgs self-coupling involving the lightest Higgs state
(hereafter $\lambda_{hhh}$) is allowed to survive. Such a circumstance
would
select the production of Higgs boson pairs as the only possible channel for
the identification of a non-SM Higgs structure. Recent studies 
at the LHC (and prospects for the luminosity 
upgrade SLHC~\cite{SLHC}) are reported in~\cite{MMPPP}, where 
the case of IMH boson pairs generated via 
WBF, associate production with $t\bar t$ pairs and with $W$/$Z$, 
each decaying into $b \bar b$ pairs, is considered in detail. 
The reason to exploit these 
non-leading production channels is due to the additional triggers available
in each case, with respect to the leading $gg\to hh$ mode, which could 
give an handle to significantly cut the background processes. 
A critical comparison 
between a complete (model-dependent) 2HDM calculation and a 
model-independent prescription, where only the 
$\lambda_{hhh}$ parameter is rescaled, shows that 
non trivial regions of the generic 2HDM parameter 
space are allowed, giving, as a signal, an excess in the $X4b$ signatures 
considered (in particular $4b + 2$ forward jets). 
A detailed account on existing studies on Higgs-pair production 
can be found in~\cite{MMPPP}. 

\section{The decoupling limit of 2HDMs}
\label{sec:2hdm}
A study of the decoupling limit of 2HDMs has recently been 
presented in~\cite{2HDM} (see also \cite{Fawzi}), where 
the general expressions for the
spectrum and couplings of a generic, non-CP violating, 2HDM are
derived.  
The deviations from the
decoupling limit are proportional to $\epsilon=\hat\lambda v^2/m_A^2$
and $\epsilon^2$ in the case of the couplings to fermions and gauge bosons,  
whereas the self-coupling is proportional to
$\epsilon\hat\lambda/\lambda$, with $\lambda$ and $\hat\lambda$ being
function of the 2HDM parameters $\lambda_i$ ($i=1,...7$) 
of Ref.~\cite{2HDM}. 
The possibility that the ratio $\hat\lambda/\lambda$ be large 
allows for the triple-Higgs self-coupling to remain large even
when the other couplings are converging to their SM values. We analyzed
this possibility~\cite{MMPPP} by implementing the exact couplings of 
a generic 2HDM as discussed in~\cite{2HDM} and scanning the parameter 
space in the range $1<\tan\beta<50$, 
$-4\pi < \lambda_i < 4 \pi$ for all couplings
$\lambda_i$ defined in~\cite{2HDM},
$i=1,...7$. Our general scan was subject to the constraints of 
tree-level unitarity~\cite{2HDM} and to the requirement that the 
couplings $g^2_{hVV}$, $g^2_{htt}$ and $g^2_{hbb}$ differ from the SM 
values by no more than 30\%, 30\% and 70\%, respectively. These values 
reflect the measurement accuracies expected after 300 \ifb\ 
of accumulated LHC luminosity~\cite{AtlasCMS}\footnote{Also a more optimistic 
scenario of 20\%, 20\% and 30\% of measurement 
accuracies~\cite{leshouches03} has been investigated, yielding
conclusions similar to the ones outlined below, though over a restricted
parameter range (see also next footnote).}. 
The distribution of $r=
\lambda_{hhh}/\lambda_{hhh}^{(0)}$ for the three Higgs mass
values of 120, 130 and 140 GeV in the general case ($\lambda_{6,7}\neq 0$)
is shown in Fig.~\ref{fig:ldist}, where the scan assumed 
equiprobable input values for all 2HDM inputs. In addition we require 
no visibility at $3\sigma$ level of the heavy neutral states $H$ and $A$, 
resulting in the mass constraints $m_H \gsim 300$~GeV 
and $m_A \gsim 250$~GeV. 
Operatively, we define as {\em decoupling region} of the 2HDM 
the configurations of 2HDM parameters which survive 
the above constraints. 
The scan of all $\lambda_i$ leads to models with values of
$r$ in the ranges $-8\lsim r\lsim 36$, $-7\lsim r\lsim 35$ and 
$-6\lsim r\lsim 34$ for $m_h = 120$, $130$ and $140$~GeV, 
respectively\footnote{In the 
case of the more optimistic uncertainty scenario we get 
$-3.5\lsim r\lsim 18$, 
$-3\lsim r\lsim 17$ and 
$-2\lsim r\lsim 16$ for $m_h = 120$, $130$ and $140$~GeV, respectively.}. 
\begin{figure}
\epsfxsize120pt
\figurebox{180pt}{200pt}{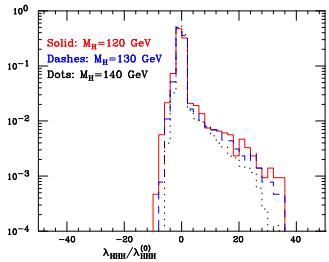}
\caption{
Distribution
of the values of 
$r=\lambdahhh/\lambdahhhzero$ 
in the scans of the 2HDM parameters space, 
for three values of the lightest Higgs boson mass. Normalization
is to unity.}
\label{fig:ldist}
\vspace*{-0.75truecm}
\end{figure}
We refer to~\cite{MMPPP} for all the details on the used selection 
criteria and on the signal-to-background analysis performed
with the ALPGEN event generator~\cite{ALPGEN} 
as well as a modification of 
{\small HDECAY} \cite{HDECAY} and of the program described in \cite{BRs}.
In order to have model-independent predictions,  
the effects of an anomalous \lambdahhh\ 
coupling have been estimated by simply rescaling the value of
$\lambdahhhzero$ to a generic $ \lambdahhh$. Over the range allowed 
by our scan, cross section enhancements by up to 
two orders of magnitude can be obtained.
Exclusion limits (at $95\%$ CL) and signal evidence (at $3\sigma$) for
anomalous triple-Higgs self-couplings by combining all channels are given
in Tab.~\ref{tab:signcombined}. 
The by far most sensitive mode is WBF, 
with top-quark($W/Z$) associate production being of some relevance
only for small Higgs masses and positive 
$\lambda_{hhh}/\lambda_{hhh}^{(0)}$ values(no relevance whatsoever). 
\begin{table*}
\begin{center}
\begin{tabular}{|l|rr|rr|rr|}
\hline
\mh~(GeV) & $120$ & & $130$ & & $140$ & \\ 
\hline
LHC, 95\% CL & $-1.7$ & $5.2$ & $-2.4$ & $5.5$ & $-3.9$ & $6.9$  \\ 
SLHC, 95\% CL & $-0.1$& $3.5$ & $-0.5$ & $3.6$ & $-1.5$ & $4.5$  \\    %
\hline
LHC, 3$\sigma$  & $-2.6$ & $6.1$ & $-3.4$ & $6.5$ & $-5.3$ & $8.3$  \\ 
SLHC, 3$\sigma$ & $-0.6$ & $4.0$ & $-1.1$ & $4.2$ & $-2.3$ & $5.2$  \\ %
\hline
\end{tabular}
\caption{Constraints on the ratio 
$\lambda_{hhh}/\lambda_{hhh}^{(0)}$ using the 
channels $X4b$. The results are almost
completely driven by WBF. 
In the top box, the two values in each entry
correspond to $r_{\rm min}$, $r_{\rm max}$, where $r<r_{\rm min}$ and
$r>r_{\rm max}$ define the range which can be excluded at 95\% CL 
(first row)    
or probed at the 3$\sigma$ level
(second row),  
at both the LHC and SLHC. The number of events corresponding 
to $3 \sigma$ significance are about 130, 
110 and 100 for $m_h= 120$, 
$130$ and $140$~GeV respectively at the SLHC.}\label{tab:signcombined}
\end{center}
\vspace*{-0.75truecm}
\end{table*}
Notice that the described rescaling is not a gauge invariant 
operation, because diagrams involving other genuine 2HDM fields (namely,
$H, A$ and $H^\pm$)
in the Lagrangian, 
as well the rescaling of the Higgs-to-SM-particle couplings to the 2HDM values,
are neglected. Therefore, one might suspect that the cross sections
 are anomalously enhanced by unitarity 
violation effects even in decoupling regime.
To investigate this possibility we have re-computed the 
WBF cross section, including 
the full set of 2HDM diagrams,
for 436 points uniformly distributed 
over the 2HDM parameter space,
fulfilling the decoupling conditions described previously, and  
with $r$ in the ranges accessible at the SLHC. 
As a results of this exercise, we did find points, for any value of $r$, 
where the 2HDM cross section agrees almost exactly with the 
model-independent approximation (see Fig.~\ref{fig:plot1}).
\begin{figure}
\begin{center}
\epsfig{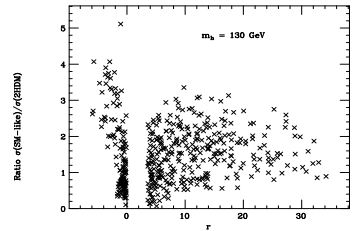}
\caption{Ratio between the SM cross section with rescaled value
of $\lambdahhhzero$ and the full 2HDM prediction, as described in the text. 
Only the region of $r$ accessible at the SLHC 
(Tab.~\ref{tab:signcombined}) is shown. 
}
\label{fig:plot1}
\end{center}
\vspace*{-0.5truecm}
\end{figure}
As a further check, we also verified, using the same points,
that the correct decoupling limit is recovered for large values
of the mass of the heavy Higgs states (see Fig.~\ref{fig:plot6}).
\begin{figure}
\begin{center}
\epsfig{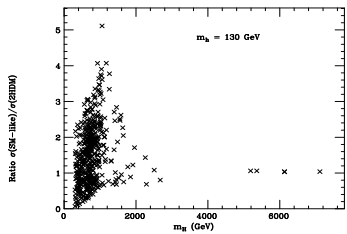}
\caption{As the previous figure, as a function of $m_H$.}
\label{fig:plot6}
\end{center}
\vspace*{-0.5truecm}
\end{figure}
Since the calculation is now gauge invariant, 
the existence of points with ratio $\sigma({\rm SM-like})/
\sigma({\rm 2HDM}) \approx 1$ demonstrates that the approximation 
of neglecting additional graphs and coupling modifications does not
lead to an artificial enhancement of the sensitivity to $\lambda_{hhh}$.
Furthermore, for $\approx70 \%$  of the points in Fig.~\ref{fig:plot1}
the cross section is reproduced by the above described
approximation within a factor two. 
Finally, for a small, but sizable, portion of points
the cross section is substantially underestimated while for even fewer
points is grossly overestimated.
We also tried to understand the origin of 
such discrepancies. Firstly, the most noticeable ones are due to the 
BR$(h \to b\bar b)$, as our definition of decoupling region 
allows for rather sizable deviations of the latter
 from the SM value and, since
we are looking for two Higgs bosons decaying into $b \bar b$, 
this BR enters quadratically into our predictions. Secondly, also the
diagrams  proportional to 
$g_{hVV}$ ($V= W,Z$) can give an important contribution: 
although the decoupling limit strongly constraints deviations of $g_{hVV}$ 
from the SM value, the consequences of the latter
can be enhanced by large destructive interferences. 
Once these two effects are accounted for, the overall agreement is fairly good:
the contribution of the SM diagrams to the total cross section is correct 
within a 20\% accuracy, except for a small fraction of points where the
production rate is strongly underestimated.
We have finally
verified that the latter effect is due to neglecting diagrams where 
a light 2HDM Higgs pair is produced via the decay of 
the heavy 2HDM Higgs boson, $H\to hh$.
Notice that the effect of this additional contribution reinforces our conclusion that 
the signal is detectable, actually suggesting that a search is possible 
even in regions where $\lambda_{hhh}$ exhibits deviations from 
the SM value smaller than those in Tab.~\ref{tab:signcombined}.

\section{Summary}
The main conclusion of our analysis is that in a large
portion of the explored 2HDM region the production of an IMH pair
will be detectable at the (S)LHC through the study of the reactions 
of the type $pp \to Xhh \to X4b$. We emphasize that, 
due to the decoupling constraints, this anomalous enhancement
of the $b\bar b b\bar b$ signal will be, in the decoupling scenario, 
the only accessible signature of departure from the SM at the LHC.
In closing, we note that our results are not confined
to the 2HDM in the decoupling limit but are model independent, thereby
being applicable to other Higgs sectors displaying a similar 
decoupling 
behavior between the lightest CP-even Higgs state and the heavier ones.

\section*{Acknowledgments}
FP thanks the BSM Session conveners for the invitation. 
MM, SM and RP acknowledge the financial support of the European Union
under contract HPRN-CT-2000-00149. RP also acknowledges the financial support
of MECD under contract SAB2002-0207. ADP thanks the Physics Department 
of the University of Bari for its kind hospitality.


\begin{thebibliography}{99}

{\small

\bibitem{AtlasCMS} 
S. Asai et al.,
SN-ATLAS-2003-024; 
M. D\"uehrssen, ATL-PHYS-2003-030; 
S. Abdullin et al., CMS Note 2003/33.

\bibitem{SLHC} 
F. Gianotti, M.L. Mangano and T. Virdee (conveners),
{\tt hep-ph/0204087}.

\bibitem{MMPPP}
M.~Moretti, S.~Moretti, F.~Piccinini, R.~Pittau and A.D.~Polosa, 
{\tt hep-ph/0410334}

\bibitem{2HDM} J.F. Gunion and H.E. Haber,
Phys. Rev. {\bf D67} (2003) 075019.

\bibitem{Fawzi} F. Boudjema and A.V. Semenov, Phys. Rev. {\bf D66}
(2002) 095007; I.F.~Ginzburg and M.~Krawczyk, {\tt hep-ph/0408011};
S. Kanemura, S. Kiyoura, Y. Okada, E. Senaha and C.-P. Yuan,
Phys. Lett. {\bf B558} (2003) 157; {\tt hep-ph/0209326}; 
S.~Kanemura, Y.~Okada, E.~Senaha and C.-P.~Yuan, {\tt hep-ph/0408364}; 
S.~Kanemura, Y.~Okada, E.~Senaha, {\tt hep-ph/0410048}.


\bibitem{leshouches03} 
M. D\"uehrssen
et al., 
{\tt hep-ph/0406323}
and in 
{\tt hep-ph/0406152}.

\bibitem{ALPGEN} M.L. Mangano, M. Moretti, F. Piccinini, R. Pittau
and A.D. Polosa,
JHEP {\bf 0307}  (2003) 001.


\bibitem{HDECAY} 
A.~Djouadi, J.~Kalinowski and M.~Spira,
Comput.\ Phys.\ Commun.\  {\bf 108} (1998) 56.

\bibitem{BRs}
S.~Moretti and W.J.~Stirling,
Phys.\ Lett. {\bf B347} (1995) 291,
Erratum, ibidem {\bf B366} (1996) 451.

}
\end{thebibliography}
\end{document}